\documentclass[preprint,showpacs,amsmath,amssymb,floatfix]{revtex4}

\usepackage{graphicx}
\usepackage{dcolumn}
\usepackage{bm}
\usepackage{slashbox}

\begin{document}

\title{Heuristic Segmentation of a Nonstationary Time Series}

\author{Kensuke Fukuda$^{1,2}$, H. Eugene Stanley$^1$, 
and Lu\'{\i}s A. Nunes Amaral$^3$}

\affiliation{ $^1$ Center for Polymer Studies and Department of Physics,
Boston University, Boston, Massachusetts 02215 \\ 
$^2$ NTT Network Innovation Laboratories, 180-8585, Tokyo, Japan\\
$^3$ Department of Chemical Engineering, 
Northwestern University, Evanston, Illinois 60208-3120
}


\begin{abstract}

Many phenomena, both natural and human-influenced, give rise to
signals whose statistical properties change under time translation,
i.e., are nonstationary. For some practical purposes, a nonstationary
time series can be seen as a concatenation of stationary segments.
However, the exact segmentation of a nonstationary time series is a
hard computational problem which cannot be solved exactly by existing
methods.  For this reason, heuristic methods have been proposed.
Using one such method, it has been reported that for several cases of
interest---e.g., heart beat data and Internet traffic
fluctuations---the distribution of durations of these stationary
segments decays with a power law tail.  A potential technical
difficulty that has not been thoroughly investigated is that a
nonstationary time series with a (scale-free) power law distribution
of stationary segments is harder to segment than other nonstationary
time series because of the wider range of possible segment sizes.
Here, we investigate the validity of a heuristic segmentation
algorithm recently proposed by Bernaola-Galv\'an {\it et
al.}~[Phys.~Rev.~Lett.~{\bf 87}, 168105 (2001)] by systematically
analyzing surrogate time series with different statistical properties.
We find that if a given nonstationary time series has stationary
periods whose size is distributed as a power law, the algorithm can
split the time series into a set of stationary segments with the
correct statistical properties. We also find that the estimated power
law exponent of the distribution of stationary-segment sizes is
affected by (i) the minimum segment size, and (ii) the ratio
$R\equiv\sigma_\epsilon/\sigma_{\bar{x}}$ where $\sigma_{\bar{x}}$ is
the standard deviation of the mean values of the segments, and
$\sigma_\epsilon$ is the standard deviation of the fluctuations within
a segment.  Furthermore, we determine that the performance of the
algorithm is generally not affected by uncorrelated noise spikes or by
{\it weak\/} long-range temporal correlations of the fluctuations within
segments. 
\end{abstract}


\maketitle

\section{Introduction}

A stationary time series has statistical properties that do not change
under time translation \cite{Stratonovich81}.  Interestingly, the time
series that arise in a large number of phenomena in a broad range of
areas---including physiologic systems, economic systems, vehicle
traffic systems, and the Internet
\cite{Ivanov,Bunde00,Goldberger02,Stanley02,Musha76,Leland95,Paxson96,Crovella97,Takayasu,Abbot99}---are
nonstationary.  Thus nonstationarity is a property common to both
natural and human-influenced phenomena.  For this reason, the
statistical characterization of the nonstationarities in real-world
time series is an important topic in many fields of research and
numerous methods of characterizing nonstationary time series have been
proposed \cite{Method}.

One useful approach to quantifying a nonstationary time series is to
view it as consisting of a number of time segments that are themselves
stationary.  The statistical properties of the segments (i) can help
us better understand the overall nonstationarity of the time series
and (ii) yield practical applications. For example, developing control
algorithms for Internet traffic will be easier if we understand the
statistical properties of these stationary segments---a behavior that
directly corresponds to the ``coarser flow rate'' of the current
network traffic.

In general, it is impossible to obtain the exact segmentation of a
nonstationary time series because of the complexity of the
calculation.  An exact segmentation algorithm requires a computation
time that scales as $O(N^N)$, where $N$ is the number of points in the
time series.  Hence, such an algorithm is not practical.  For this
reason, the segmentation of a real-world time series must accomplish a
trade-off between the complexity of the calculation and the desired
precision of the result.

Bernaola-Galv\'an and co-workers recently proposed a heuristic
segmentation algorithm \cite{Segment} designed to characterize the
stationary durations of heart beat time series. In this algorithm
\cite{Segment}, the calculation cost is reduced by iteratively
attempting to segment the time series into only {\it two\/} segments.
A decision to cut the times series is made by evaluating a modified
Student's $t$-test for the data in the two segments.  

The application of this segmentation algorithm reveals that the
distribution of the stationary durations in heart beat time series
decays as a power law \cite{Segment}.  Intriguingly, a recent analysis
of Internet traffic uncovered that the distribution of stationary
durations in the fluctuation of the traffic flow density also follows
a power law dependence \cite{Fukuda01}. Because these signals have
their origin in such diverse contexts, these findings suggest that the
power law decay of the distribution of the stationary period may be a
common occurrence for complex time series.  Thus, the correct
implementation and interpretation of the results obtained by the
segmentation algorithm is essential in understanding the dynamics of
the system.  In fact, there are many implementation issues concerning
the segmentation algorithm of Ref.~\cite{Segment} that have not yet
been addressed explicitly in the literature, especially those
concerning the proper estimation of the value of the power-law tail's
exponent in the cumulative distribution of stationary durations.

In this paper we systematically analyze different types of surrogate
time series to determine the scope of validity of the segmentation
algorithm of Bernaola-Galv\'an and co-workers \cite{Segment}.  In
Section II, we briefly explain the segmentation algorithm.  In Section
III, we present results for the dependence of the exponent of the
power law tail on the minimum size of the segments in the distribution
of the stationary durations.  In Section IV, we consider the effect of
the amplitude of the noise and the presence of spike-type noise.  In
Section V we consider long-ranged temporally-correlated noise.
Finally, in Section VI, we summarize our findings.

\section{Implementing the Segmentation Algorithm}

\subsection{The algorithm}

To divide a nonstationary time series into stationary segments
\cite{Segment}, we move a sliding pointer from left to right along the
time series and, at each position of the pointer, compute the mean of
the subset of the signal to the left of the pointer $\mu_{\rm left}$
and to the right $\mu_{\rm right}$.  For two samples of Gaussian
distributed random numbers, the statistical significance of the
difference between the means of the two samples, $\mu_{1}$ and
$\mu_{1}$, is given by the Student's $t$-test statistic
\cite{prob_intro}
\begin{equation}
t\equiv \left|\frac{\mu_{1} - \mu_{2}}{S_D}\right|, 
\end{equation}
where 
\begin{equation}
S_D = \left(\frac{(N_{1} - 1) s_{1}^2 + (N_{2} - 1)
s_{2}^2}{N_{1} + N_{2} - 2}\right)^{1/2}
\left(\frac{1}{N_{1}} + \frac{1}{N_{2}}\right)^{1/2}
\end{equation}
is the pooled variance \cite{NRC}, $s_{1}$, $s_{2}$ are the standard
deviations of the two samples, and $N_{1}$ and $N_{2}$ are the number
of points in the two samples.  

Moving the pointer along our time series from left to right, we
calculate $t$ as a function of the position in the time series. We use the
statistic $t$ to quantify the difference between the means of the
left-side and right-side time series. Larger $t$ means that the values
of the mean of both time series are more likely to be significantly
different, making point $t_{\rm max}$, with the largest value of $t$, a
good candidate as a cut point.

We then calculate the statistical significance $P(t_{\rm max})$.  Note
that $P(t_{\rm max})$ is not the standard Student's $t$-test because we
are not comparing independent samples.  $P(t_{\rm max})$ is
numerically approximated as
\begin{equation}
P(t_{\rm max}) \approx \{1 - I_{[\nu/(\nu+t_{\rm max}^2)]}\times
(\delta\nu, \delta)\}^{\gamma}, 
\end{equation}
where $\gamma = 4.19\ln N - 11.54$ and $\delta = 0.40$ are obtained from 
Monte Carlo simulations \cite{Segment}, $N$ is the size of the time
series to be split, $\nu = N - 2$, and $I_x(a, b)$ is the incomplete
beta function.  

If the difference in mean is not statistically significant---i.e., if
is smaller than a threshold (typically set to 0.95)---then the time
series is not cut.  If the difference in means between the left and
right part of the time series is statistically significant, then the
time series is cut into two segments as long as the means of the two
new segments are significantly different from the means of the
adjacent segments.  If the time series is cut, we continue iterating
the above procedure recursively on each segment until the obtained
significance value is smaller than the threshold, or the size of the
obtained segments is smaller than a minimum size $\ell_0$.  We will
see that the value of $\ell_0$ is one of the parameters controlling
the accuracy of the algorithm.

\subsection{Surrogate Time Series}

To investigate the validity of the algorithm, we analyze surrogate time
series $x(t)$ generated by linking segments with different means.  As
described in Section~I, the cumulative distribution of the stationary
durations for some real-world time series is characterized by a power
law decay in the tail, so the probability of finding a segment of size
larger than $m$, i.e., the cumulative distribution of segment sizes in
our time series, is
\begin{equation}
P(>m) = \frac{\gamma+1}{m_0^{\gamma+1}} m^{-\gamma}, \ \ \ \mbox{for} \ m \gg m_0,  
\label{eq:e3}
\end{equation}
where $m_0$ is the minimum size of a segment.
\begin{figure}[htbp]
\centering
\includegraphics*[width=.5\textwidth]{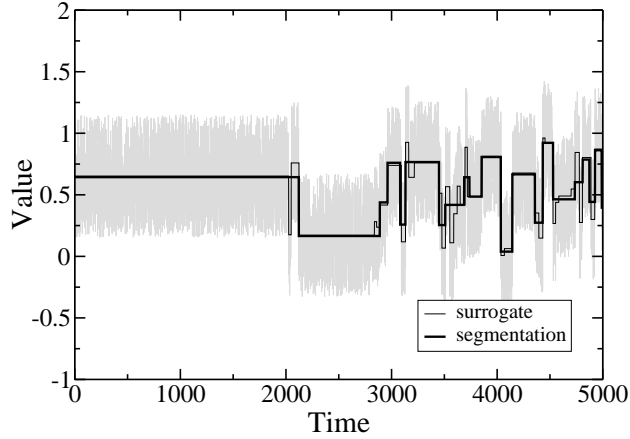}
  \caption{\label{fig:1} Surrogate time series constructed according
    to the procedure described in II.B with parameters $m_0 = 20$,
    $\gamma = 1.0$, and $R=1$.  The time series is drawn in light
    gray, and the stationary segments are represented by a dark gray
    line.  The black line displays the output of the segmentation
    algorithm for $\ell_0 = 50$.  It is visually apparent the black
    line provides a adequate coarse-grained description of the
    surrogate time series. Note also that the algorithm cannot extract
    the smallest segments because of the restrictions on resolution
    for $m < \ell_0$.  }
\end{figure}

We generate time series with a power law distribution of segment sizes
by the following procedure:

\begin{enumerate}

\item draw from the interval [$m_0$,+$\infty$] a sequence of segment
lengths \{$m_i$\} with distribution given by Eq.~(\ref{eq:e3});

\item draw from the interval [0,1] a sequence of mean time series
values $\bar{x}_{i}$ with uniform probability;

\item draw from the interval
[-$\sqrt{3}\sigma_\epsilon$,$\sqrt{3}\sigma_\epsilon$] a sequence of
fluctuation values $\epsilon_i(k_i)$, for $k_i=1,\cdots,m_i$, with
uniform probability.

\end{enumerate}

The resulting time series is given by
\begin{equation}
x(t) = \bar{x}_{i} + \epsilon_i(k_i)\,,
\label{eq:1}
\end{equation}
where $i$ is such that 
\begin{equation}
\sum_{j=0}^{i-1}m_j < t \le \sum_{j=0}^{i}m_j\,,
\end{equation}
and $k_i$ is such that
\begin{equation}
k_i = t - \sum_{j=0}^{i-1}m_j\,.
\end{equation}

To quantify the level of the noise, we define the ratio
\begin{equation}
R \equiv{\sigma_\epsilon\over\sigma_{\bar{x}}}, 
\end{equation}
where $\sigma_{\bar{x}}$ is the standard deviation of the mean of the
segments and $\sigma_\epsilon$ is the standard deviation of the
fluctuations within a segment.  For $\bar{x}$ uniformly distributed
in the interval [0,1], $\sigma_{\bar{x}} = 0.3$.

For each set of parameters ($R$, $\gamma$, $m_0$) we generate ten time
series, each with 50,000 data points.  Note that knowing {\it a
priori\/} the value of $m_0$ in a real-world time series is unlikely,
but in order to test the algorithm in a consistent way, we consider in
the following section $m_0 \ge 20$ because that is the resolution
limit for the algorithm (see also Appendix A).

\begin{figure}[htbp]
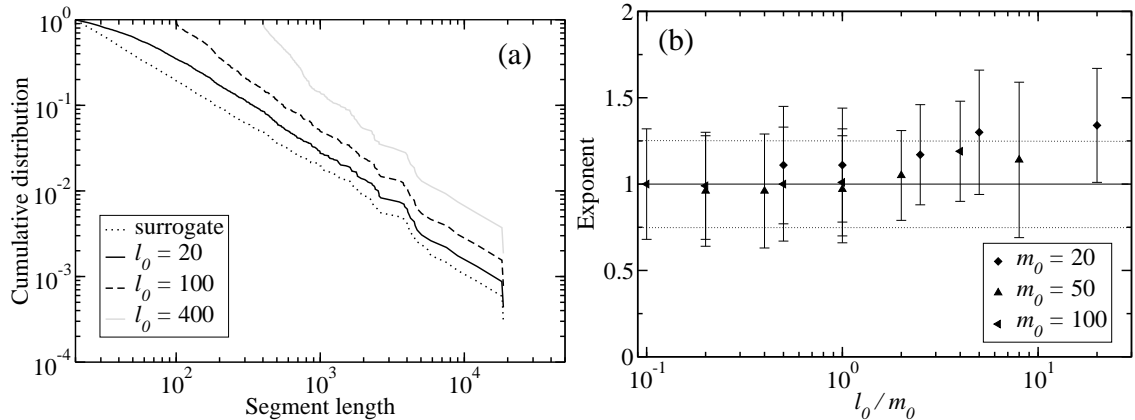

\centering
\includegraphics*[width=.45\textwidth]{fig2a.eps}
\includegraphics*[width=.45\textwidth]{fig2b.eps}
  \caption{\label{fig:2} (a) Cumulative distribution $P(\ell)$ of
    segment sizes larger than $\ell$ for surrogate time series with
    50,000 data points; $\gamma = 1.0$, $R=1$ and $m_0 = 20$.  The
    dotted line indicates the input segment size distribution for the
    surrogate time series.  The slope of the line is $1.0$ for $20 < m
    < \mbox{4,000}$.  The other curves show $P(\ell)$ for different
    values of $\ell_0$.  For $\ell_0 = 20$, the curve is well
    described in the range $100 < \ell < \mbox{4,000}$ by a power law
    with $\hat{\gamma} \approx 1.0$.  The distribution does not decay
    as a power law for $\ell < 100$ due to the fact that the
    segmentation algorithm cannot split a time series with a number of
    points insufficient to perform the Student's $t$-test. For $\ell_0
    = 400$, $P(\ell)$ decays as a power law for $\mbox{1,000} < \ell <
    \mbox{8,000}$.  The segmentation algorithm correctly splits all
    segments in the surrogate series with size $\ell > \mbox{1,000}$.
    (b) Dependence of $\hat{\gamma}$ on $\ell_0$ and $m_0$. The mean
    and the standard deviation of the exponent for the original time
    series is $1.05\pm0.24$, shown by the black solid and dotted
    lines. The error bars show the standard deviation of the estimates
    $\hat\gamma$. For $\ell_0/m_0 < 5$, we find $\hat{\gamma} \approx
    1$.  Thus, the algorithm accurately estimates exponents in this
    region.  However, the values of the exponent are close to 1.3 for
    $\ell_0/m_0 > 10$, meaning that $m_0\ll\ell_0$ leads to an
    overestimation of $\gamma$.}
\end{figure}

Figure~\ref{fig:1} displays a surrogate time series, the corresponding
stationary durations, and the result of the segmentation algorithm
with $\ell_0 = 50$ and $m_0 = 20$.  The segments obtained by the
segmentation algorithm do not exactly match the stationary segments in
the surrogate time series but the figure strongly suggests that the
algorithm provides the correct coarse-grained description of the time
series.  As expected, the segmentation algorithm cannot extract
segments with size $m < \ell_0$.

\section{Accuracy of the segmentation algorithm}

\subsection{Dependence on $\ell_0$ and $m_0$}

Figure~\ref{fig:2}(a) displays the cumulative distribution of segment
sizes, which is the probability of finding a segment with size larger
than $\ell$ for a surrogate time series split for different values 
$\ell_0$.  The cumulative distributions of the size of the stationary
segments cut by the segmentation algorithm for surrogate time series
are well fit by a power law decay.  For all cases, we find
\begin{equation}
P_{\ell_0, m_0}(>\ell) \sim \ell^{-\hat{\gamma}(\ell_0,m_0)} 
\end{equation}
with the same exponent value $\hat{\gamma} \approx 1.0$ \cite{GAMMA},
indicating that the segmentation algorithm splits the nonstationary
time series into segments with the correct asymptotic statistical
properties.  However, the range of scales for which we observe a power
law decay with $\hat\gamma \approx \gamma$ depends strongly on the
selection of $\ell_0$.

\begin{table*}[htbp]
\centering
  \caption{\label{tab:pow-1.0} Estimated exponent $\hat{\gamma}$, as
  defined by Eq.~(5) for $\gamma = 1.0$.  The mean and the standard
  deviation of the exponents are calculated for the ranges indicated
  inside parenthesis using Eq.~(\ref{eq:sgamma}).  The column labeled
  ``input'' presents exponent estimates obtained from the segment
  lengths used to generate the surrogate time series.  We find
  $\hat{\gamma}\approx \gamma = 1.0$ for $m_0 < 5 \ \ell_ 0$.}
\begin{tabular}{|c|c|c|c|c|c||c|}\hline
\backslashbox{$m_0$}{$\ell_0$} & 10 & 20 & 50 & 100 & 400 & input\\\hline\hline
20 & $1.1\pm0.3$  & $1.1\pm0.3$  & $1.2\pm0.3$  & $1.3\pm0.4$  & $1.3\pm0.3$  & $1.1\pm0.3$\\
& ($\ell>20$) & ($\ell>20$) & ($\ell>50$) & ($\ell>110$) & ($\ell>1000$) & \\\hline
50 & $1.0\pm0.3$  & $1.0\pm0.3$  & $1.0\pm0.3$  & $1.1\pm0.3$  & $1.1\pm0.5$  & $1.0\pm0.2$\\
& ($\ell>50$) & ($\ell>50$) & ($\ell>100$) & ($\ell>110$) & ($\ell>1000$) & \\\hline
100 & $1.0\pm0.3$ & $1.0\pm0.3$ & $1.0\pm0.3$ & $1.0\pm0.3$ & $1.2\pm0.3$ & $1.1\pm0.3$\\
& ($\ell>100$) & ($\ell>100$) & ($\ell>100$) & ($\ell>100$) & ($\ell>1000$) & \\\hline
\end{tabular}
\end{table*}

For $\ell$ greater than about $5 \ell_0$, the tails of the
distributions are close to power law decays with $\hat{\gamma}\approx
\gamma = 1.0$.  Moreover, all $P(\ell)$ follow the original curve for
$\ell > \mbox{1,000}$, i.e., the algorithm correctly identifies the
large segments independently of the selection of $\ell_0$.  For $\ell
< 5 \ell_0$, the distributions do not follow power law decays.  The
origin of this behavior lies in the fact that (i) for $\ell_0 = 20 =
O(m_0)$, there are not enough data points to reliably perform the
Student's $t$-test, so one cannot reasonably expect any statistical
procedure to be able to extract those short segments, and (ii) for
$\ell_0 \gg m_0$, one fails to extract many of the stationary
segments.  The reason for the latter is that the value of $\ell_0$ is
in this case considerably larger than the size of the shortest
segments in the time series, so the algorithm is forced to merge a
number of short segments into longer ones with size greater than
$\ell_0$.  This process gives rise to an excess of segments with sizes
between $\ell_0$ and $2\ell_0$.

Table \ref{tab:pow-1.0} shows the mean and standard deviation of the
estimated exponent value $\hat{\gamma}$ calculated from surrogate time
series for several values of $m_0$ and $\ell_0$ (see Appendix A for
details on how to to estimate $\hat{\gamma}$).  Our results indicate
that $\hat{\gamma}$ depends on both $m_0$ and $\ell_0$: If $\ell_0 \gg
m_0$, $\hat{\gamma}$ overestimates $\gamma$, while if $\ell_0 \approx
0(m_0)$, the algorithm correctly estimates the value of the exponent
$\gamma$; cf. Fig.~\ref{fig:2}(b).

\begin{figure}[htbp]
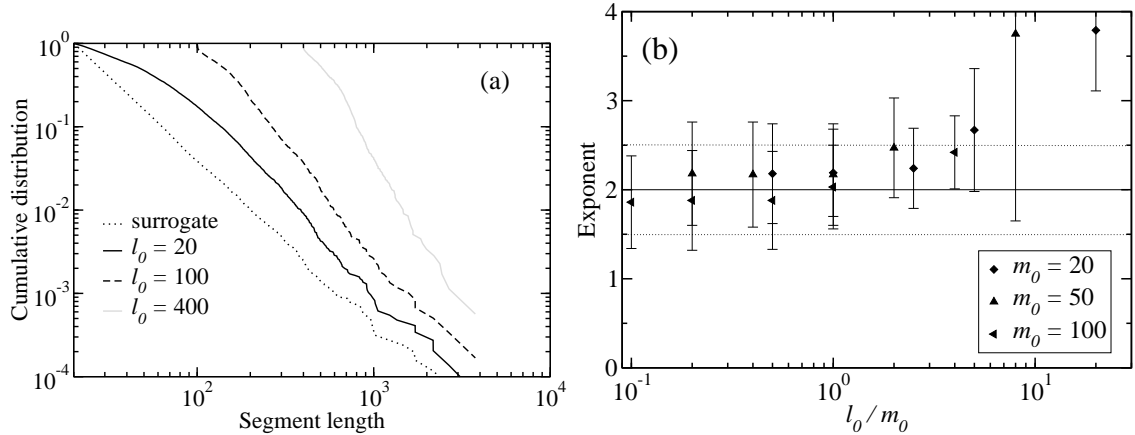

\centering
\includegraphics*[width=.45\textwidth]{fig3a.eps}
\includegraphics*[width=.45\textwidth]{fig3b.eps}
   \caption{\label{fig:3} (a) Cumulative distribution of segment sizes
     for $\gamma = 2.0$, $R=1$ and $m_0 = 20$.  For $\ell_0 = 20$ and
     $100 < \ell < \mbox{1,000}$, the exponent $\hat{\gamma}$ of the
     power law is close to $\gamma$.  For $\ell_0 = 100$, we also
     $\hat{\gamma} \approx \gamma$. However, for $\ell_0 = 400$, the
     algorithm fails to split the time series correctly for $\ell <
     \mbox{1,000}$.  Moreover, note that even though the exponent
     estimate is correct, the algorithm yields segments that are
     longer than the ones in the surrogate time series.
     (b) Dependence of $\hat{\gamma}$ on $\ell_0$ and $m_0$ for
     $\gamma = 2.0$.  For $\ell_0/m_0 < 4$, the algorithm yields
     segments with the correct statistical properties.}
\end{figure}

\subsection{Dependence on $\gamma$}

Next, we focus on the dependency of the accuracy of the segmentation
algorithm on the value of $\gamma$.  Figure~\ref{fig:3}(a) displays
the cumulative distribution of segment sizes for surrogate time series
generated with $\gamma = 2.0$.  A challenge for the segmentation
algorithm is that Eq.~(\ref{eq:e3}) indicates that the probability of
finding segments with size shorter than $4m_0$ is 90\% for $\gamma =
2.0$, which can lead to the ``aggregation'' of several consecutive
segments of small size into a single longer segment.

As we found for $\gamma = 1$, the tails of the distributions follow
power law decays for large $\ell$, showing that the algorithm yields
segments with the proper statistical properties.  In
Fig.~\ref{fig:3}(b), we show the dependence of $\hat{\gamma}$ on
$\ell_0$ and $m_0$ for $\gamma = 2.0$.  We find that for $\ell_0/m_0 <
4$ the algorithm extracts segments with distributions of lengths that
decay in the tail as power laws with exponents that are quite close to
$2.0$ .

\begin{table*}[htbp]
\centering
   \caption{\label{tab:pow-2.0} Estimated exponents $\hat{\gamma}$ for
     $\gamma = 2.0$. The mean and the standard deviation of the
     exponents are calculated for the ranges indicated inside
     parenthesis using Eq.~(\ref{eq:sgamma}).  The column labeled
     ``input'' presents exponent estimates obtained from the segment
     lengths used to generate the surrogate time series.  We find
     $\hat{\gamma}\approx \gamma = 2.0$ for $m_0 < 4 \ \ell_ 0$. }
\begin{tabular}{|c|c|c|c|c|c||c|}\hline
\backslashbox{$m_0$}{$\ell_0$} & 10 & 20 & 50 & 100 & 400 & input\\\hline\hline
20 & $2.2\pm0.6$  & $2.2\pm0.5$  & $2.2\pm0.5$  & $2.4\pm0.7$  & $3.8\pm0.7$  & $2.2\pm0.1$\\
& ($\ell>100$) & ($\ell>100$) & ($\ell>100$) & ($\ell>200$) & ($\ell>800$) & \\\hline
50 & $2.2\pm0.6$  & $2.2\pm0.6$  & $2.2\pm0.6$  & $2.5\pm0.6$  & $3.8\pm2.1$  & $2.1\pm0.4$\\
& ($\ell>100$) & ($\ell>100$) & ($\ell>100$) & ($\ell>200$) & ($\ell>1000$) & \\\hline
100 & $1.9\pm0.5$ & $1.9\pm0.6$ & $1.9\pm0.6$ & $2.0\pm0.5$ & $2.4\pm0.4$ & $2.0\pm0.4$\\
& ($\ell>100$) & ($\ell>100$) & ($\ell>100$) & ($\ell>200$) & ($\ell>1000$) & \\\hline
\end{tabular}
\end{table*}

\begin{table*}[htbp]
\centering
   \caption{\label{tab:pow-3.0} Estimated exponents $\hat{\gamma}$ for
     $\gamma = 3.0$. The mean and the standard deviation of the
     exponents are calculated for the ranges indicated inside
     parenthesis using Eq.~(\ref{eq:sgamma}).  The column labeled
     ``input'' presents exponent estimates obtained from the segment
     lengths used to generate the surrogate time series.  For this
     value of $\gamma$, one finds that a small $m_0$ leads to a clear
     under-estimation of the $\gamma$.  Note that the standard
     deviation of $\hat{\gamma}$ becomes larger, indicating the
     difficulty in obtaining $\hat{\gamma}$ accurately.  Also
     noteworthy is the fact that because $\gamma$ is so large, the
     range of segment lengths $m$ drawn becomes much reduced.  This
     implies that if one sets $\ell_0 = 400$ one is unable to properly
     estimate $\gamma$.}
\begin{tabular}{|c|c|c|c|c|c||c|}\hline
\backslashbox{$m_0$}{$\ell_0$} & 10 & 20 & 50 & 100 & 400 & input\\\hline\hline
20 & $2.4\pm0.5$ & $2.3\pm0.5$  & $2.5\pm0.5$ & $2.7\pm0.5$  & $4.3\pm1.4$  & $3.1\pm0.5$\\
& ($\ell>100$) & ($\ell>100$) & ($\ell>100$) & ($\ell>200$) & ($\ell>1000$) & \\\hline
50 & $3.0\pm0.8$  & $3.0\pm0.8$  & $3.0\pm0.9$  & $3.4\pm0.6$  & $5.9\pm1.3$  & $3.0\pm0.6$\\
& ($\ell>100$) & ($\ell>100$) & ($\ell>100$) & ($\ell>200$) & ($\ell>1000$) & \\\hline
100 & $2.8\pm0.9$ & $2.8\pm0.8$ & $2.8\pm0.9$ & $2.8\pm0.8$ & $5.2\pm1.7$ & $2.9\pm0.7$\\
& ($\ell>200$) & ($\ell>200$) & ($\ell>200$) & ($\ell>200$) & ($\ell>1000$) & \\\hline
\end{tabular}
\end{table*}

In Tables~\ref{tab:pow-2.0} and \ref{tab:pow-3.0}, we report the
values of $\hat{\gamma}$ for $\gamma = 2.0$ and $\gamma = 3.0$,
respectively.  We find that for small $m_0$ and large $\gamma$, one
over-estimates $\gamma$. Thus, we surmise that for $\gamma > 3.0$, it
becomes impractical to estimate $\gamma$ accurately, except for
extremely long time series.  This fact is not as serious a limitation
as one may think because for large $\gamma$ it is always difficult to
judge whether a distribution decays in the tail as an exponential or
as a power law with a large exponent.

\section{Robustness of the algorithm with regard to noise}

\subsection{Amplitude of fluctuations around a segment's mean}

Another factor that may affect the performance of the segmentation
algorithm of Bernaola-Galv\'an and co-workers is the amplitude of the
fluctuations within a segment.  It is plausible that greater noise
amplitudes will increase the difficulty in identifying the boundaries
of neighboring segments.  Thus, we next analyze the effect of the
amplitude of the noise for surrogate time series.  

\begin{figure*}[htbp]
\centering
\includegraphics*[width=.9\textwidth]{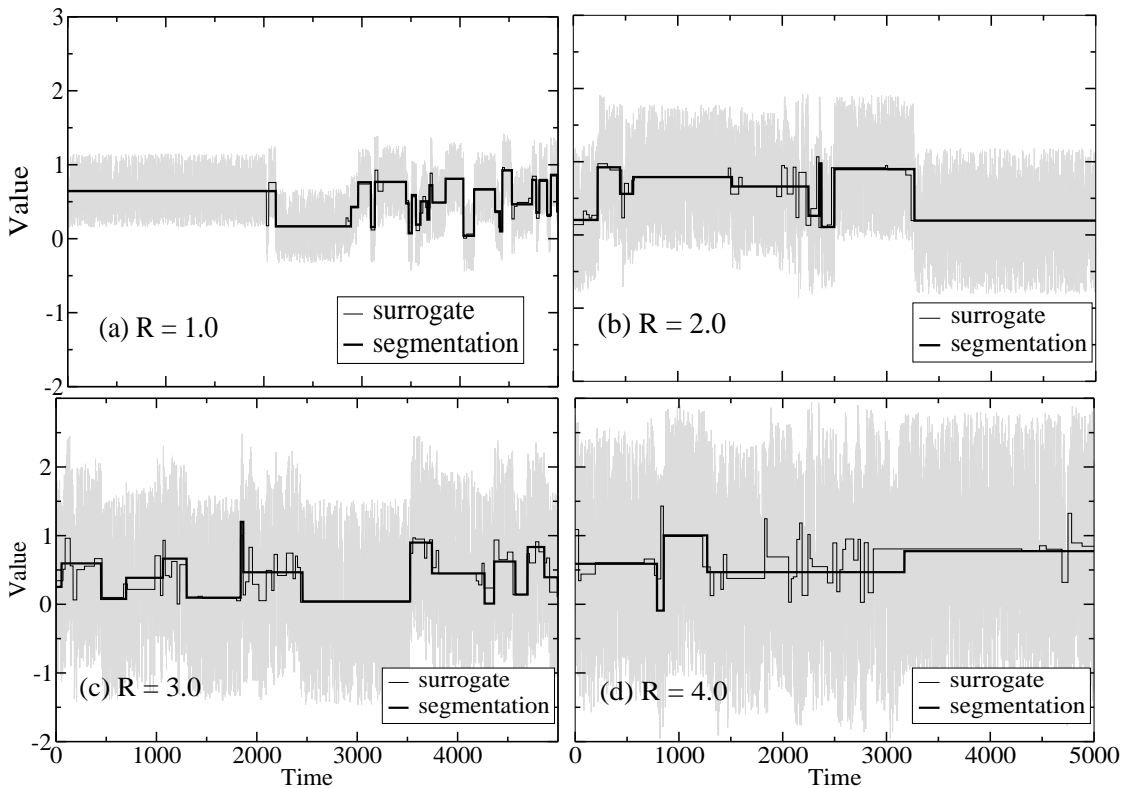}
  \caption{\label{fig:raw-noise}Surrogate time series for different
  amplitudes of the fluctuations within the segments.  The surrogate
  time series were generated with $m_0 = 20$, $\ell_0 = 20$, $\gamma =
  1.0$, and (a) R = 1.0, (b) R = 2.0, (c) R = 3.0, and (d) R = 4.0.
  It is visually apparent that the segmentation algorithm yields
  longer segments as $R$ increases.  This fact arises from the fact
  that the statistical test cannot distinguish two neighboring
  segments with close---relative to $\sigma_\epsilon$---means.}
\end{figure*}

Figure~\ref{fig:raw-noise} demonstrates that for large R, the
segmentation algorithm yields few short segments.  This result arises
from the concatenation of neighboring segments with means that become
statistically indistinguishable due to the large value of
$\sigma_\epsilon$.

\begin{figure*}[htbp]
\centering
\includegraphics*[width=.9\textwidth]{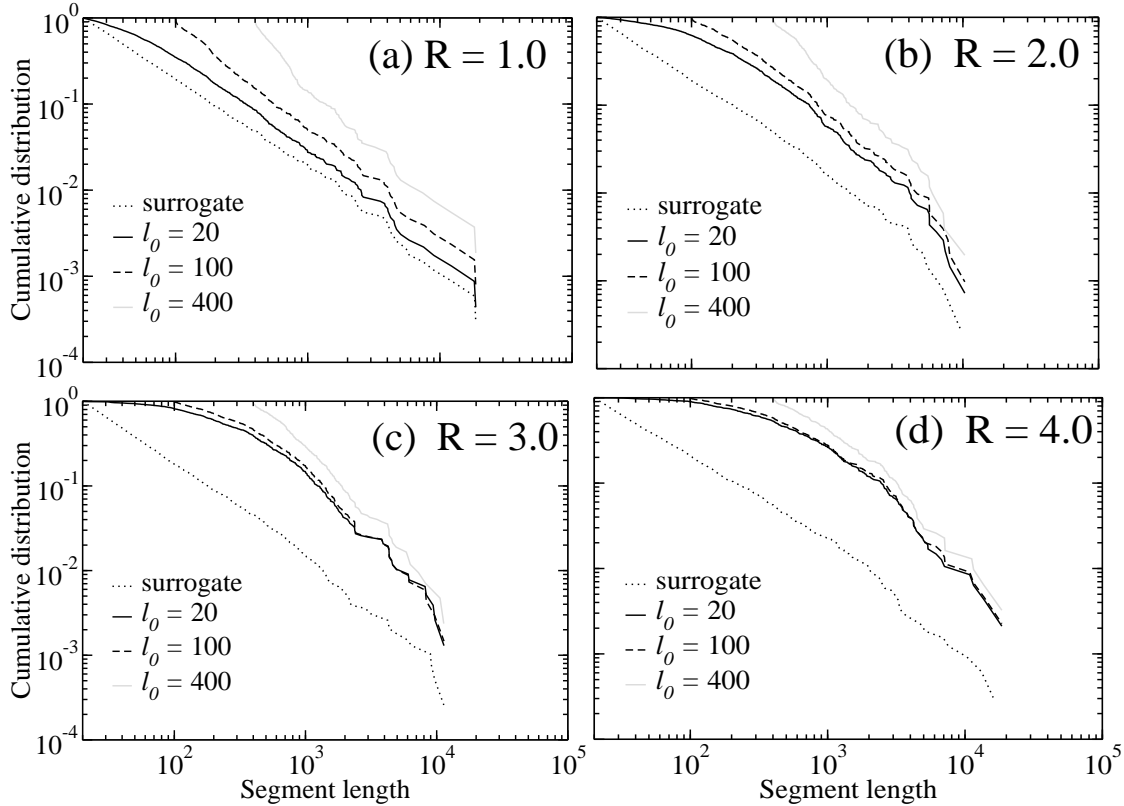}
  \caption{\label{fig:4} Cumulative distribution of segment size
  obtained with the segmentation algorithm for different amplitudes of
  the noise.  As in Fig.~\protect\ref{fig:1}, the time series
  analyzed were generated for $m_0 = 20$, $\gamma = 1.0$, and (a) $R =
  1.0$, (b) $R = 2.0$, (c) $R = 3.0$, and (d) $R = 4.0$.  For $R \le
  3$, the tails of the distributions decay as power laws.  For $R =
  4.0$, it is difficult to discriminate whether our the tails of
  distribution conform to exponential or power law decays.  Note that
  as $R$ increases, the dependence of the functional form of the
  distributions on $\ell_0$ decreases appreciable.}
\end{figure*}

We show in Fig.~\ref{fig:4} the cumulative distributions of segment
sizes for $\gamma = 1.0$, $m_0 = 20$, and for different values of $R$.
For large $\ell$ and $R \le 3$, we find $\hat{\gamma} \approx \gamma$,
while for $R=4$ the algorithm becomes ineffective at extracting the
stationary segments in the time series.  It is visually apparent that
for $R = 4.0$ the fluctuations within a segment are so much larger
than the jumps between the means of the stationary segments that the
segmentation becomes totally unable to parse the different segments;
cf.  Fig~\ref{fig:raw-noise}d.

\begin{figure*}[htbp]
\centering
\includegraphics*[width=.9\textwidth]{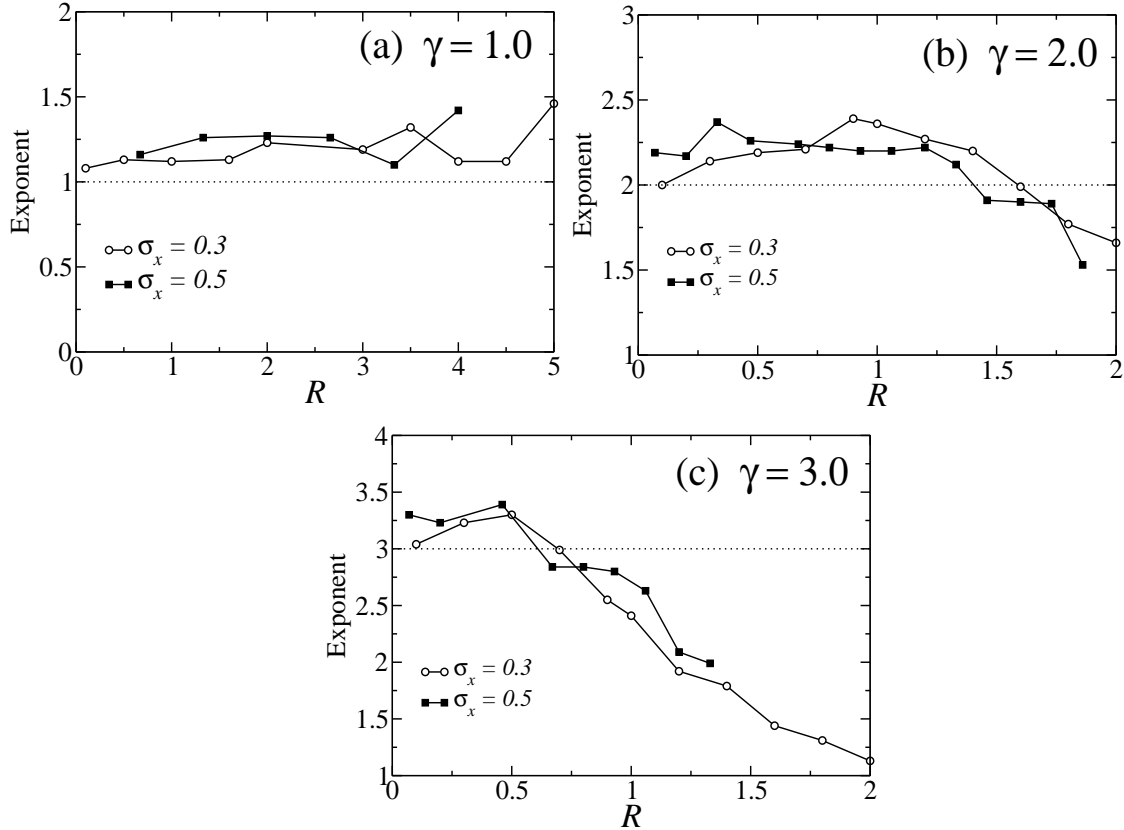}
   \caption{\label{fig:5} Dependence of $\hat{\gamma}$ on the ratio
      $R=\sigma_{\epsilon}/\sigma_{\bar{x}}$ for $m_0 = 20$, $\ell_0 =
      20$ and three distinct values of $\gamma$: (a) $\gamma = 1.0$,
      (b) $\gamma = 2.0$, and (c) $\gamma = 3.0$. The different curves
      in each plot correspond to different values of $\sigma_\mu$. For
      $\gamma = 1.0$, we find $\hat\gamma = 1.2$ for $0 < R < 4.0$,
      indicating that the algorithm is robust against increases in
      $R$. For $\gamma > 1.0$, we find that the impact of an
      increasing $R$ on the performance of the algorithm becomes more
      and more marked. Specifically, for $\gamma = 2.0$, we find
      $\hat\gamma \approx \gamma$ for $R < 1.5$, while for $\gamma =
      3.0$, we find $\hat\gamma \approx \gamma$ for $R < 0.6$. }
\end{figure*}

In Fig.~\ref{fig:5}, we show $\hat{\gamma}$ for different values of
$R$.  For $\gamma = 1.0$, we estimate $\hat{\gamma} \approx \gamma$
for $0 < R < 4$. In contrast, for $\gamma =2.0$, $\hat{\gamma} \approx
\gamma$ only for $0 < R < 1.5$.  When $R$ increases, the segmentation
algorithm is unable to cut the segments because the greater amplitude
of the fluctuations inside a segment decreases the significance of the
differences between regions of the time series.  This effect yields
very large segments, which results in very small estimates of
$\hat{\gamma}$.  This effect is even stronger for $\gamma = 3.0$, for
which we find $\hat{\gamma} \approx \gamma$ only for $0 < R < 0.6$.

\subsection{Spike noise}

\begin{figure*}[htbp]
\centering \includegraphics*[width=.9\textwidth]{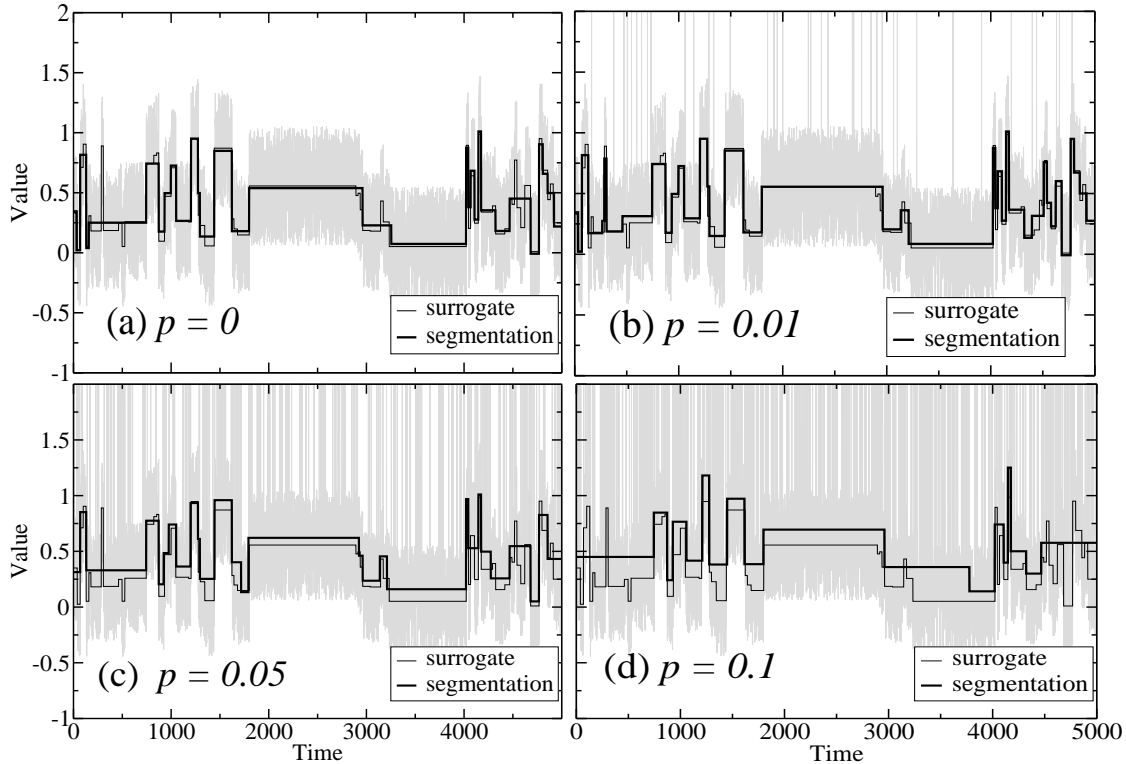}
  \caption{\label{fig:raw-pulse}Surrogate time series with
  uncorrelated spike noise for $m_0 = 20$, $\gamma_\epsilon = 0.3$, 
  $\ell_0 = 20$, $\gamma =  1.0$ and (a) $p = 0$, (b) $p = 0.01$, (c) $p = 0.05$, and (d) $p =
  0.1$.  For all values of $p$ considered, the application of the
  segmentation algorithm yields segments that, in a coarse-grained
  way, match well the segments in the surrogate data the.}
\end{figure*}

\begin{figure*}[htbp]
\centering
\includegraphics*[width=.9\textwidth]{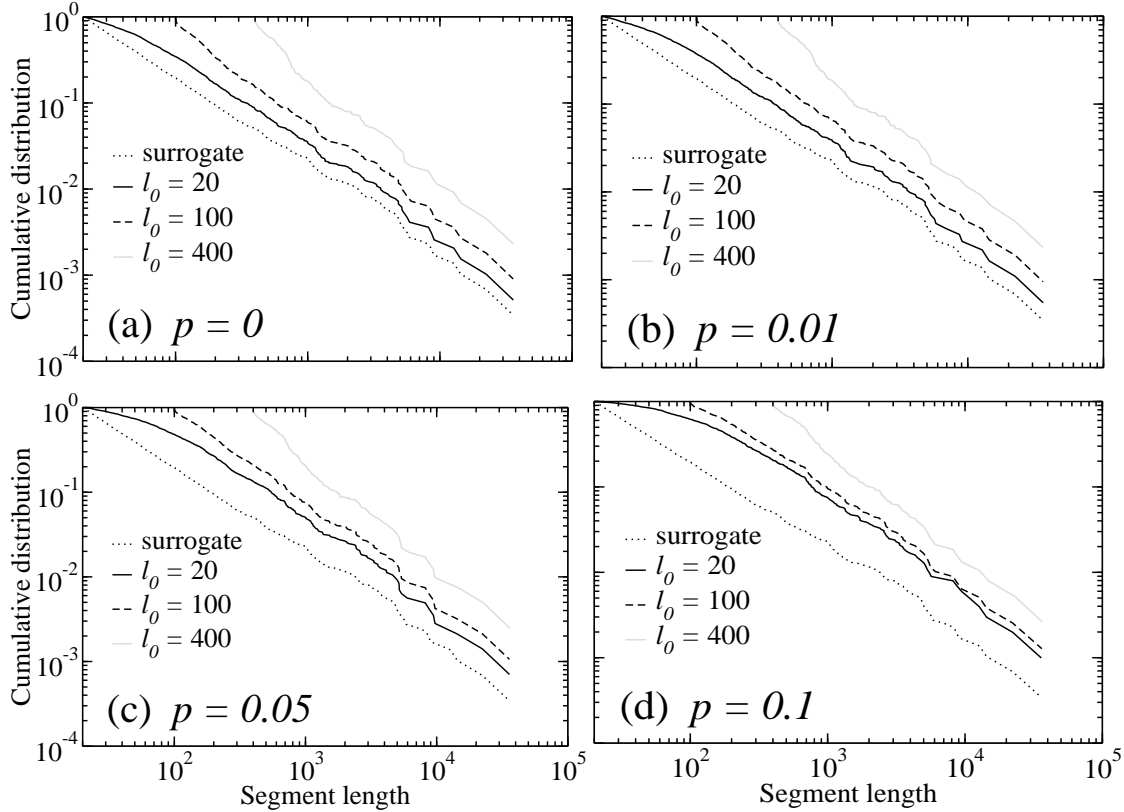}
   \caption{\label{fig:6} Cumulative distribution of segment sizes for
     surrogate time series with $\gamma = 1.0$, $\sigma_\epsilon = 0.3$, 
     $m_0 = 20$, and for  different densities $p$ of the spike noise: (a) $p = 0.0$, (b) $p
     = 0.01$, (c) $p = 0.05$, and (d) $p = 0.1$.  The value of $x(t)$
     is set at 2.0 for a spike.  It is visually apparent that even for
     $p = 0.1$ all the distributions decay as power laws for $\ell >
     200$ and that the slopes in the log-log plots are similar to the
     slope of the input distributions, i.e., the segmentation
     algorithm yields segments with the correct statistical properties
     even in the presence of strong spike noise.}
\end{figure*}

Next we analyze the effect of spike noise on the performance of the
segmentation algorithm.  We generate surrogate time series as before
and then for each $t$ make, with probability $p$, $x(t) = 2$.  The
effect of this procedure is illustrated in Fig.\ref{fig:raw-pulse} for
four distinct values of $p$.  The figure also suggests that the
segmentation algorithm yields a good coarse-grained description of the
surrogate time series for $p$ as large as 0.1.  This result suggests
that the algorithm is robust to the existence of uncorrelated spike
noise in the data (Fig.~\ref{fig:6}).

\section{Correlated noise}
In this Section, we investigate the effect of long-range correlations
in the fluctuations around the segment's mean on the performance of
the segmentation algorithm. This study is particularly important because
real-world time series often display long-range power law decaying
correlations.  
\subsection{Segmentation of correlated noise with no segments}
\begin{figure}[htbp]
\centering
  \includegraphics*[width=.95\textwidth]{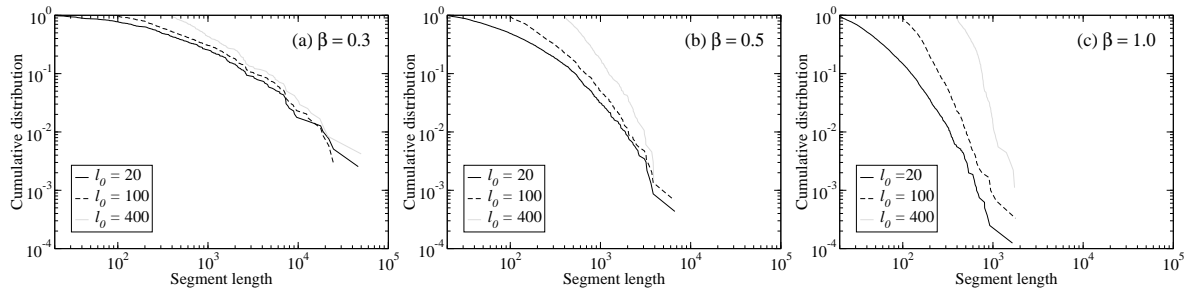}
\vspace{-0.0cm}
\caption{\label{fig:seg_corr_onesegment} Segmentation of Guassian distributed
long-range correlated noise with different exponent of the power law 
in the power spectrum density for the noise; 
(a) $\beta = 0.3$, (b) $\beta = 0.5$, and (c) $\beta = 1.0$. 
 We generate time series with 60,000
data points according to the modified Fourier filtering method (FFM)
\cite{Makse96}.  
It is hard for all the curves to determine the functional form; 
a power law or an exponential distribution.
There exists only one long segment for $\beta = 0$. 
Then, for larger $\beta$, the segmentation algorithm splits the time series 
into many smaller pieces of stationary segments. 
This is because that 
the time series with larger $\beta$ has a higher probability 
that a part of the time series is nonstationary. 
However, a quantitative relationship between $\hat{\gamma}$ and 
$\beta$ is not clear for all the case of the correlated noise with no segments. 
}
\end{figure}

We generate a temporally-correlated noise 
whose power spectrum decays as $S(f) \sim f^{-\beta}$ 
\cite{Makse96}.  The surrogate time series consists of
60,000 points, with mean 0. 
Figure~\ref{fig:seg_corr_onesegment} displays the cumulative
distribution of segment lengths.  The curves show the different noise
correlations: $\beta = 0.3$,  0.5, and 1.0 ($1/f$).  
Note that we have confirmed only one long segment for $\beta = 0.0$, as one expects.
On the other hand,  for small $\beta$, we still observe a longer segment, although 
the curves decay rapidly for large $\beta$. 
Thus, the segmentation algorithm divides 
a correlated noise for large $\beta$  into many small pieces of stationary segments, 
because the given time series is already nonstationary  for this case. 
Also, it is hard to determine the functional form of the plots for all cases. 
Even if the plots are assumed to follow power laws, 
the relationship between $\beta$ and $\hat{\gamma}$  is quantitatively obscure. 


\subsection{Segmentation of correlated noise with segments}

\begin{figure*}[htbp]
\centering
\includegraphics*[width=.95\textwidth]{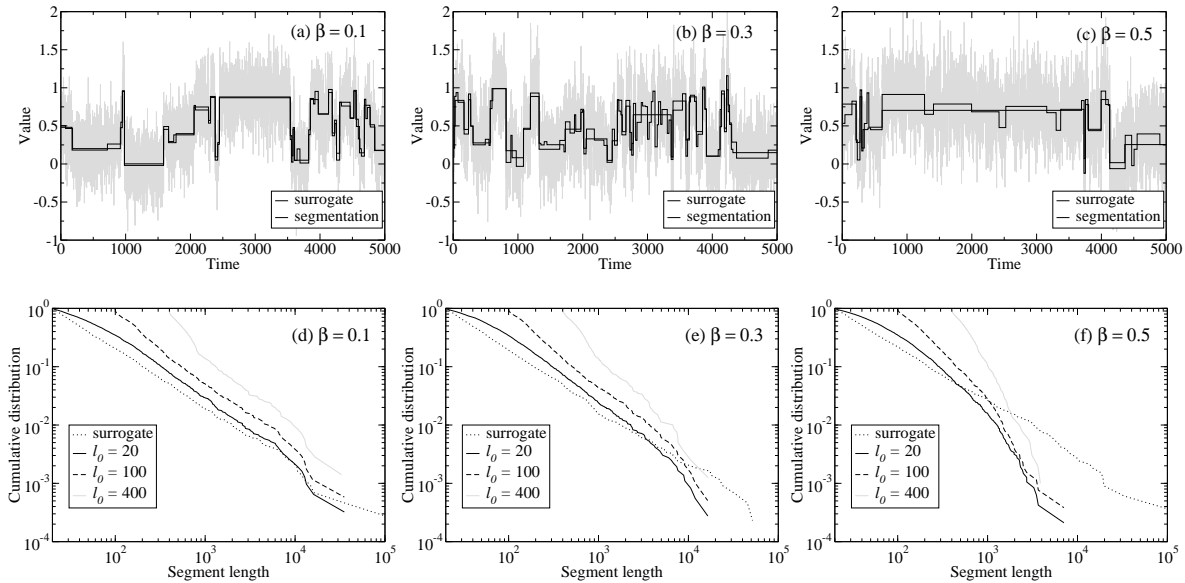}
  \caption{\label{fig:betaraw} Surrogate time series with the
  long-range correlated noise and the result of the segmentation for
  (a) $\beta = 0.1$, (b) $\beta = 0.3$, and (c) $\beta = 0.5$.  The
  time series was concatenated by segments with the long-range
  correlated noise generated by the modified Fourier filtering method
  (FFM) \cite{Makse96} for $\gamma=1.0$, $m_0 = 20$, $\ell_0 = 20$,
  and $R=1$.  For $\beta = 0.1$ and $0.3$, the results of the
  segmentation algorithm resemble the segments in the surrogate time
  series.  However, for $\beta = 0.5$, it is visually apparent that
  long segment are cut multiple times, indicating that the algorithm
  judges the noise within a segment as a nonstationary time series.
  This is because that for large $\beta$, the correlated noise in a
  segment cannot be stationary.  The same is not true for short
  segments, which are identified correctly under the above conditions.
  Cumulative distributions of the segument size for (d) $\beta = 0.1$,
  (e) $\beta = 0.3$, and (f) $\beta = 0.5$.  The distributions confirm
  the visual impression obtained form (a)-(c).  In summary, the
  distributions decay faster than a power law when $\beta > 0.3$,
  although they still follow a power law dependence for smaller
  $\beta$. }
\end{figure*}

Here, in order to show the ability of the identification of  
segments, we analyze surrogate time series concatenating segments 
with long-range correlated random variables
whose power spectrum is followed by a power law with an exponent $\beta$.
We show in Figs.~\ref{fig:betaraw}(a)-(c) typical surrogate time series
generated with $\gamma = 1.0$, $m_0 = 20$, $R = 1.0$, $\ell_0 = 20$, 
and different temporal correlations: (a) $\beta = 0.1$, 
(b) $\beta = 0.3$, and (c) $\beta = 0.5$.  
The figure suggests that the segmentation
algorithm can correctly parse the short segments but that long
segments get cut multiple times, especially for $\beta = 0.5$. 
This result is to be expected
because the strong correlations in the noise lead to marked changes in
the mean.

In Figs.~\ref{fig:betaraw}(d)-(f), we show the cumulative
distributions of segment sizes for $\beta = 0.1,\ 0.3, \mbox{and }
0.5$.  The data confirm quantitatively the visual impression gained
from Figs.~\ref{fig:betaraw}(a)-(c), i.e., that longer segments get
cut multiple times.  In particular, for $\beta = 0.5$, the
distributions clearly deviate from the power law, independent of the
selection of $\ell_0$.  However, this fact should not be seen as a
shortcoming of the algorithm; for large $\beta$, correlated noise for
a long segment is already nonstationary, so that the algorithm is
indeed cutting a nonstationary signal into stationary durations.


\section{Discussion}

In this paper we analyzed nonstationary surrogate time series with
different statistical properties in order to investigate the validity
of the segmentation algorithm of Bernaola-Galv\'an and co-workers
\cite{Segment}.  Our results demonstrate that this heuristic
segmentation algorithm can be extremely effective in determining the
stationary regions in a time series provided that a few conditions are
fulfilled: First, one must have enough data points in the time series
to yield a large number of segment lengths, otherwise one will not be
able to reach the aymptotic regime of the tail of the distribution of
segment sizes.  

Second, the ratio of the amplitude of the fluctuations within a
segment to the typical jump between the means of the stationary
segments must be relatively small (less than about $0.6$) in order for
one to trust the output of the segmentation algorithm.  This concern
contrast with the case of spike noise in the data which affects the
performance of the segmentation algorithm only weakly.

Finally, if there are long-range temporal correlations of the
fluctuations around the mean of the segment, then the segmentation
algorithm correctly cuts the time series into the stationary segments
for small $\beta$.  However, for $\beta > 0.3$, the fluctuations
inside long segments become nonstationary, which results in the
algorithm detecting many ``stationary'' durations inside these long
segments.

Our analysis provides a number of clear guidelines for using
effectively the segmentation algorithm of Bernaola-Galv\'an {\it et
al.}  \cite{Segment}:
 
\begin{enumerate}

\item One must perform the segmentation for a number of different
values of $\ell_0$ in order to identify the region for which the tails
of the distributions of segment sizes reach the asymptotic scaling
behavior. (Note: If $\hat{\gamma}$ is large, than the estimation error
can be quite considerable, especially if $m_0$ is small.)

\item One must calculate the ratio between the standard deviation of
the mean value of the segment and the standard deviation of the
fluctuations within a segment after performing the segmentation.  If
the $R$ than 0.6, then there is the possibility that $\hat{\gamma}$ is
considerably under-estimating the true value of $\gamma$.

\end{enumerate}

\subsubsection*{Acknowledgments}

We thank S. Havlin, P. Ch. Ivanov, and especially P. Bernaola-Galv\'an
for discussions.  We thank NIH/NCRR (P41 RR13622) and NSF for support.
L.A.N.A. acknowledges a Searle Leadership Fund Award for support.

\appendix

\begin{figure}[htbp]
\centering
\includegraphics*[width=.45\textwidth]{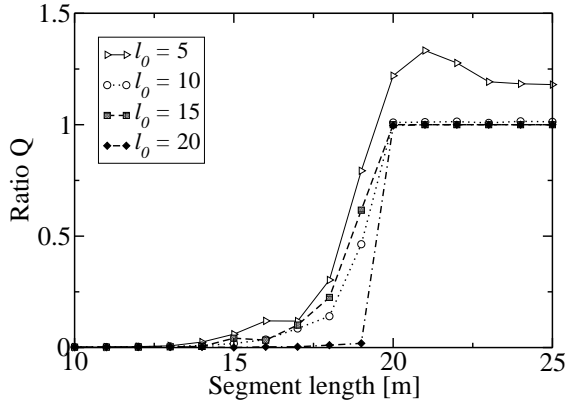}
  \caption{\label{fig:fixed} Algorithm's resolution with regard to
   $\ell_0$ and $m$. We generate surrogate time series are constructed
   by the alternated concatenation of two types segments with fixed
   size $m_0$: one with mean 0.0 and standard dewviation 0.29 and
   another with mean 1.0 and standard deviations 0.29.  We plot $Q$,
   which is defined in Eq.~(\ref{eq:q}), as a function of $m$ for
   different values of $\ell_0$.  Our results indicate that the
   segmentation algorithm is not able to extract the stationary
   segments for $m<20$ or $\ell_0 < 10$.  Note also that $\ell_0 \le
   m$ for correct segmentation to occur.}
\end{figure}

\section{Performance of the algorithm for fixed segment sizes}

As a special case, we analyze the time series whose stationary
duration is fixed in order to discuss the minimum resolution of the
segment algorithm. We concatenate segments with constant lenght $m$
with alternating means of 0.0 and 1.0.  We then add fluctuations to
those segments with a standard deviation of 0.29.  We define the
fraction of successfully split segments
\begin{equation}
Q\equiv\frac{\mbox{Number of segments correctly identified 
by the algorithm}}{\mbox{Number of
segments in surrogate data}}\,,
\label{eq:q}
\end{equation}
where $Q = 1.0$ corresponds to perfect segmentmentation.

We plot in Fig.~\ref{fig:fixed} $Q$ as a function of $m$ for different
values of $\ell_0$ .  For $m < 20$, the segmentation algorithm does
not yield the correct segments in the surrogate time series even
though the segment's means are quite different.  This result suggests
that the resolution of the segmentation algorithm is $\approx 20$.  We
also find that for $\ell_0 = 5$, the algorithm splits the time series
into too many segments.  This result suggests that for optimal
performance $\ell_0 \ge 10$.

\section{Estimation of $\hat{\gamma}$}

\begin{figure*}[htbp]
\centering
\includegraphics*[width=.9\textwidth]{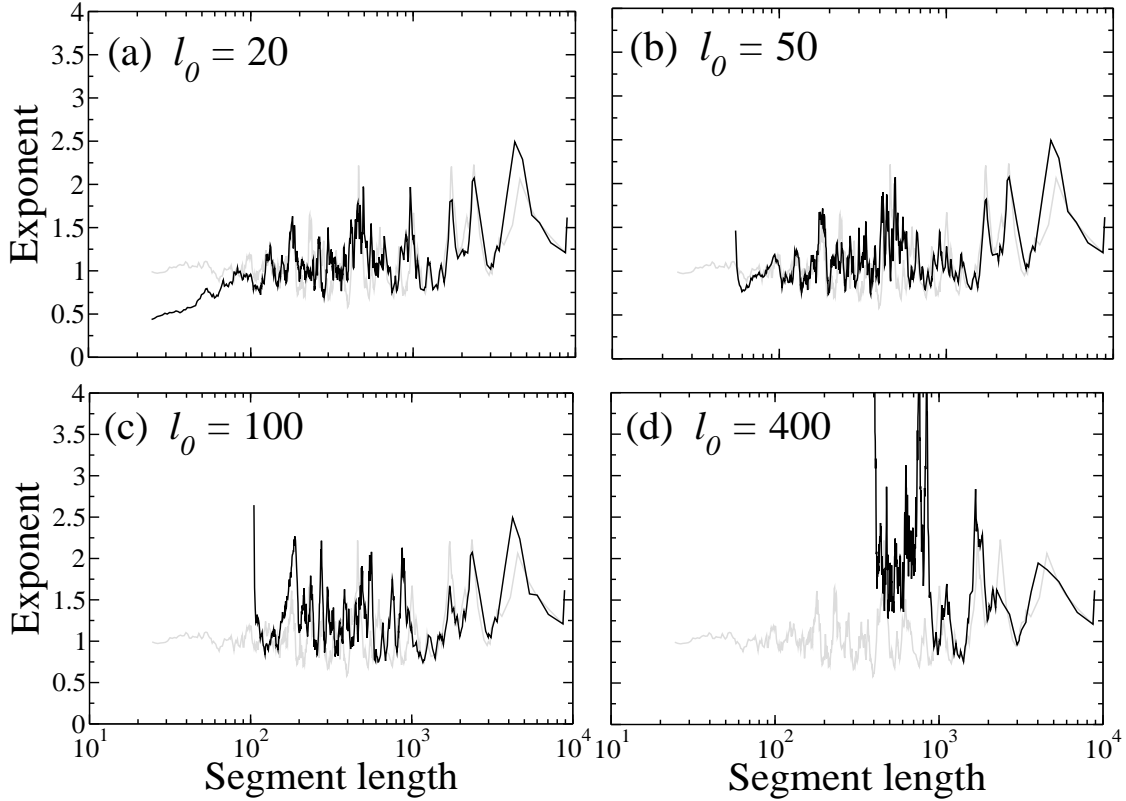}
  \caption{\label{fig:exp-base} The behavior of the exponents
    $\hat{\gamma}(\ell_n)$ of the power law in segment size
    distribution for different $\ell_0$; (a) $\ell_0 = 20$, (b)
    $\ell_0 = 50$, (c) $\ell_0 = 100$, and (d) $\ell_0 = 400$. $m_0 =
    20$, $\gamma = 1.0$. The exponent of the power law is calculated by
    Eq.~(\ref{eq:sgamma}), and plotted by black lines in the figure.
    The dotted gray lines also indicate the exponent of the surrogate
    time series for comparison.  For $\ell_0 = 20$, the observed curve
    resembles the original one for $\ell > 100$ and stays around 1.0,
    though the curve indicates an exponential decay for $\ell < 100$.
    For $\ell_0 = 100$ and $400$, the segmentation results can follow
    the original ones for larger $\ell$; however, the estimated values
    of the exponent are somewhat overestimated for smaller $\ell$}
\end{figure*}

To estimate the exponent $\hat{\gamma}$ of the power laws, we calculate
the power law exponent at a segment size $\ell_n$ from a small region
around $\ell_n$
\begin{equation}
\hat{\gamma}(\ell_n) = \frac{\log(P(\ell_{n+\tau})) - \log(P(\ell_{n}))}{
\log(\ell_{n+\tau}) - \log(\ell_{n})},
\label{eq:sgamma}
\end{equation}
where we set the region to $\tau = 5$.  Figure~\ref{fig:exp-base} shows
the behavior of the power law exponent over $\ell$ corresponding to
Fig.~\ref{fig:2}(a); the black lines are the behavior of the exponent at
segment size $\ell$, and the broken lines are the behavior of the
exponent of the power law in the distribution of the surrogate time
series for comparison.

The range of $\ell_n$ for the calculation is between a starting point
(where the exponential decay disappears) and 10,000.  For $\ell_0 =
20$, the value of the exponent decreases for $\ell < 100$, which
corresponds to the exponential decay as shown in Fig.~\ref{fig:2}(a).
However, the value of the exponent fluctuates around 1.0 for $\ell >
100$, indicating that the algorithm can reproduce the correct
statistical behavior.  For $\ell_0 = 50$, the value of the exponent is
close to 1.0 for $50 < \ell < \mbox{10,000}$.  Moreover, for $\ell_0 =
100$ and $400$, the curves decay quickly for the smaller $\ell$, and
the value of the exponent tends to be overestimated.


\begin{thebibliography}{99}

\bibitem{Stratonovich81} R. L. Stratonovich, {\it Topics in the Theory
of Random Noise, Vol.~1\/} (Gordon and Breach, New York, 1981).

\bibitem{Ivanov}
P. Ch. Ivanov {\it et al.}, 
Nature {\bf 383}, 323 (1996); 
P. Ch. Ivanov {\it et al.}, 
Nature {\bf 391}, 461 (1999); 

\bibitem{Bunde00}
A. Bunde {\it et al.}, 
Phys. Rev. Lett. {\bf 85}, 3736 (2000).

\bibitem{Goldberger02}
A. L. Goldberger {\it et al.}, 
Proc. Nat. Acad. Sci. USA {\bf 99 Supp. 1}, 2466 (2002).

\bibitem{Stanley02}
H. E. Stanley, {\it et al.},
Proc. Nat. Acad. Sci. USA {\bf 99 Supp. 1}, 2561 (2002).

\bibitem{Musha76}
T. Musha and H. Higuchi, 
Jour. Appl. Phys. {\bf 15}, 1271 (1976).

\bibitem{Leland95}
W. Leland {\it et al.}, 
Trans. Net. {\bf 2}, 1 (1995).

\bibitem{Paxson96}
V. Paxson and S. Floyd, 
Trans. Net. {\bf 3}, 226 (1996).

\bibitem{Crovella97}
M. Crovella {\it et al.}, 
Trans. Net. {\bf 5}, 835 (1997). 

\bibitem{Takayasu}
M. Takayasu {\it et al.}, 
Physica A {\bf 233}, 924 (1996);
M. Takayasu {\it et al.},
Physica A {\bf 277}, 248 (2001). 

\bibitem{Abbot99}
D. Abbott and L. B. Kish (eds), 
{\it Unsolved Problems of Noise\/} (Melville, New York, 1999).

\bibitem{Method} C-K. Peng, {\it et al.}, Chaos {\bf 5}, 82 (1995);
Z. R. Struzik, Fractals {\bf 8}, 163 (2000).

\bibitem{Segment} P. Bernaola-Galv\'an {\it et al.}, Phys. Rev. Lett. {\bf
87}, 16 (2001).

\bibitem{Fukuda01} K. Fukuda {\it et al.}, in preparation.

\bibitem{prob_intro} W. Feller, {\it An Introduction to Probability
Theory and Its Application, 2nd Ed., vol.1\/} (Willey, New York,
1971).

\bibitem{NRC} W. H. Press {\it et al.}, {\it Numerical Recipes in C\/}
(Cambridge University Press, Cambridge, 1988).

\bibitem{GAMMA} We use the notation $\hat{\gamma}$ as the abbreviation
form of $\hat{\gamma}(\ell_0, m_0)$ when the context is clear.  Also,
we use $P(\ell)$ instead of $P_{\ell_0, m_0}(\ell)$.

\bibitem{Makse96} H. A. Makse {\it et al.}, Phys. Rev. E {\bf 53},
5445 (1996). 

\end{thebibliography}
\end{document}